\documentclass[aps,prl,twocolumn,groupedaddress,showpacs]{revtex4}   

\usepackage{graphicx}

\begin{document}



\title{
        Tracing the evolution of the symmetry energy of hot nuclear fragments
        from the compound nucleus towards multifragmentation

       }

\author{G. A. Souliotis$^{1}$}

\author{A. S. Botvina$^{2}$}
\author{D. V. Shetty$^{1}$}
\author{A. L. Keksis$^{1}$}
\author{M. Jandel$^{1}$}
\thanks{C-INC, Los Alamos National Laboratory, Los Alamos, NM 87545.}
\author{M. Veselsky$^{1}$}
\thanks{Institute of Physics of the Slovak Academy of Sciences, Bratislava, Slovakia.}
\author{S. J. Yennello$^{1}$}

\affiliation{$^{1}$ Cyclotron Institute, Texas A\&M University,
College Station, TX 77843}

\affiliation{$^{2}$ Institute for Nuclear Research, Russian  Academy
of Sciences, RU-117312 Moscow, Russia   }

\date{\today}


\begin{abstract}

The evolution of the symmetry energy coefficient of the binding
energy of hot fragments with increasing excitation is explored in
multifragmentation processes following heavy-ion collisions below
the Fermi energy. In this work, high-resolution mass spectrometric
data on isotopic distributions of projectile-like fragments
are systematically compared to calculations involving the
Statistical Multifragmentation Model (SMM).
Within the SMM picture, the present study suggests a gradual 
decrease of the symmetry energy coefficient of the hot primary
fragments from 25 MeV at the compound nucleus regime 
towards 15 MeV in the multifragmentation regime.     
The isotopic distributions of the hot primary fragments are found to
be very wide and extend towards the neutron drip-line.
These findings are expected to have important implications  to the
modeling of the composition and the evolution of hot and dense
astrophysical environments, such as those of core-collapse
supernova.


\end{abstract}




\maketitle


Nuclear  multifragmentation is one of the most interesting phenomena
in nuclear physics as it holds promise for understanding nuclear   
matter properties at the extreme conditions of high excitation
energy and large isospin ($N/Z$) asymmetry
\cite{MF1,MF2,MF3,MF4,MF5}. The latter, in particular, plays a     
profound role in the dynamics of various astrophysical
environments \cite{Steiner,Prakash,Woosley,ZBao-Book}. It is well  
established (e.g. \cite{MF4}) that nuclear systems with relatively
low excitation energy ($E^{*}$/$A$ $\leq$ 2 MeV) form the
traditional compound nucleus,
whereas at higher excitation energy, the hot nuclear system
expands and, subsequently, 
disassembles into  an ensemble of hot primary fragments.
This extremely complicated process, namely the multifragmentation,
occurs on a timescale of 100 fm/c (3.3$\times$10$^{-22}$ sec) during
which the system can sample a large number of configurations. For
this reason, statistical calculations (e.g., \cite{SMM1,MMMC}) have
been very successful in describing this process.

Recently, a remarkable similarity has been pointed out between the
thermodynamic conditions (temperature, density, isospin asymmetry
$N/Z$) reached in nuclear multifragmentation
and the collapse/explosion of massive stars
\cite{Botvina-plb,Botvina-prc,Botvina-epja}. This observation opens
up the possibility of applying well-established models of nuclear
reactions to describe matter distribution and evolution during
supernova explosions \cite{Botvina-plb}. 
In addition, statistical calculations suggest that in
multifragmentation \cite{Botvina-plb,Pratt} and in hot astrophysical
environments (e.g. supernova) \cite{Botvina-plb,Ishizuka03}, the
ensemble of primary fragments includes neutron-rich nuclei towards
or beyond the neutron drip-line.

The primary fragments are expected to be hot (with excitation
energies approaching 2--3 MeV/nucleon \cite{Hudan}) and, initially, in    
close proximity to neighboring fragments or nucleons. These
conditions render their properties, e.g. binding energy, different
from those of cold (ground state) isolated nuclei.  In particular,
recent studies \cite{Iglio,Shetty05,LeFevre,Henzlova} give evidence
for a significant decrease of the symmetry energy of hot primary
fragments. We recall that for a system with $N$ neutrons and $Z$
protons, the symmetry energy
can be expressed as
 $E_{\rm sym} = C_{\rm sym} (N-Z)^{2}/A $,
with $C_{\rm sym}$ the symmetry energy coefficient and $A$ the mass
number.    
In the aforementioned studies, the symmetry energy of hot fragments
from multifragmentation at high excitation energy ($E^{*}/A$ = 4--6
MeV) is found to be reduced to $C_{\rm sym} \sim $ 15 MeV or lower,
as compared to the conventional value of $C_{\rm sym} \sim $ 25 MeV
for cold nuclei. 
In the same vein, our recent work \cite{GS-csym} indicated decreased
symmetry energy in the region $E^{*}/A $ = 2--3 MeV. 
The observed reduction of the symmetry energy implies that more
neutron-rich nuclei are favored in the distribution of fragments
after the partitioning of the initial hot and expanded nuclear
system. Such unusual neutron-rich nuclei should also be quite
abundant in supernova matter \cite{Botvina-plb} and may affect the
dynamics of the explosion and the subsequent nucleosynthesis. 

For completeness, we emphasize that presently there is strong and
growing interest in the density dependence of the symmetry energy of
nuclear matter (e.g. \cite{MF1,ZBao-Book,Famiano,Shetty-prl} and
references therein). However, we should point out that at low
densities (less than half of the normal nuclear density) the uniform
nuclear matter breaks-up into fragments.
Thus, in order to describe hot fragment distributions, we should
address the problem of the symmetry energy of individual fragments
(primary fragments) embedded in a hot and dense environment.

The present Rapid Communication is intended to provide evidence for
a continuous evolution of the symmetry energy coefficient from the
compound nucleus regime ($E^{*}/A $ $\leq$ 2 MeV) towards
multifragmentation ($E^{*}/A $$\geq$ 4 MeV) by a systematic
comparison of the isotopic distributions of heavy fragments
with calculations based on the Statistical Multifragmentation Model
(SMM). Herein, fragments with $A$ $>$ 40 are referred to as heavy
residues, those with $A$ $<$ 40 as heavy IMF (intermediate mass
fragments), whereas both groups are collectively called heavy
fragments.
Moreover, in this work, the $N/Z$ distributions of hot nuclei from
the multifragmentation of neutron-rich systems are found to be very
wide involving exotic nuclei towards the neutron drip-line.


The experimental data were obtained at the Cyclotron Institute of
Texas A\&M University employing beams from the K500 Superconducting
Cyclotron. Two magnetic separators were used: the Momentum Achromat
Recoil Separator (MARS)  for the $^{86}$Kr\,(25\,MeV/nucleon) +
$^{64}$Ni,$^{124}$Sn,$^{208}$Pb reactions and the Superconducting
Solenoid Line (BigSol) for the $^{64}$Ni\,(25\,MeV/nucleon) +
$^{64}$Ni reaction.
A detailed description of the measurements is presented in
\cite{GS-csym,GSprl,GSplb}. Employing standard  
$B\rho$--$TOF$--$\Delta E$--$E$ (magnetic rigidity, time-of-flight,
energy loss and residual energy, respectively) techniques
\cite{GSplb}, the atomic number $Z$, 
the mass number $A$  and the velocity of the fragments were obtained
with high resolution. 
The measurements of  $^{86}$Kr + $^{64}$Ni, $^{124}$Sn were
performed in the angular range 2.7$^{o}$--5.4$^{o}$ 
and $B\rho$ range 1.3--2.0 T\,m. 
The measurements of the $^{86}$Kr + $^{208}$Pb were obtained in the
angular range 1.0$^{o}$--2.7$^{o}$ 
and $B\rho$ range 1.2--1.7 T\,m. 
Finally, the $^{64}$Ni + $^{64}$Ni measurements were performed with
BigSol in the angular range 1.5$^{o}$--3.0$^{o}$ and $B\rho$ range
1.1--1.6 T\,m \cite{GS-csym}.
The measured yield distributions were used to
extract the average $Z/A$ values. 
To obtain total fragment cross sections, the measured yield data
were corrected for the spectrometer acceptance (as in \cite{GSplb})
with the aid of the simulations described below.


The calculations are based on a two-stage Monte Carlo approach. The
dynamical stage of the collision was described by the deep-inelastic
transfer (DIT) code of Tassan-Got \cite{DIT} simulating stochastic
nucleon exchange in peripheral and semiperipheral collisions. This
model has been successful in describing  the isospin, excitation
energy and kinematical properties of  excited quasiprojectiles in a
variety of recent studies \cite{MV1,GSplb,MV2,GS-plb04}.
The deexcitation of the quasiprojectiles was performed with the
latest version of the Statistical Multifragmentation Model (SMM)
\cite{Botvina-epja,Botvina-markov,SMM1}, referred to as ``SMM05''
and briefly summarized below.
This  model assumes statistical equilibrium at a low-density
freeze-out stage and includes all breakup channels ranging from the
compound nucleus to vaporization (channels with only light particles
$A$$<$4), allowing  a unified description of nuclear disintegration
with increasing excitation.
In the microcanonical treatment, the statistical weight of a decay
channel is calculated as exponential of the entropy. Light fragments
with $A$$<$4 are considered as stable particles (nuclear gas) with
only translational degrees of freedom. Fragments with $A$$>$4 are
treated as heated liquid drops with free energies parameterized as a
sum of volume, surface, Coulomb and symmetry energy terms
\cite{Botvina-epja} with parameters  adopted from the
Bethe-Weizsacker mass formula.
The model generates a Markov chain of partitions (with the
Metropolis algorithm) representative of the whole partition ensemble
\cite{Botvina-markov}.
In the freeze-out configuration, the hot primary fragments are
assumed to be isolated and at normal density; they, subsequently,
propagate in their mutual Coulomb field, while undergoing binary
deexcitation via evaporation, fission or Fermi breakup \cite{SMM1}.
The Coulomb interaction energy is directly calculated for each
spatial configuration of fragments in the freeze-out volume.
Finally, the effect of the Coulomb field of the target in proximity
to the decaying quasiprojectile is included.
The symmetry energy evolution is taken into account in the mass
calculation of the primary fragments and the secondary fragments
during the deexcitation stage, as in \cite{Botvina-epja}. Below a
threshold of $E^{*}/A $=1 MeV, a smooth transition to standard
experimental masses is assumed \cite{Botvina-epja}.
Standard parameters of the SMM code are employed. A multiplicity
dependent parametrization of the free volume \cite{SMM1}
(determining the contribution of the fragment translational motion
to the partition probability \cite{SMM1}) is used, whereas, the
freeze-out volume (determining the Coulomb energy of the fragment
partition) is taken to be 6 times the nuclear volume at normal
density (as suggested by the asymptotic velocities of the observed
heavy fragments).

    \begin{figure}[h]                                        
    \includegraphics[width=0.47 \textwidth, height=0.65 \textheight ]{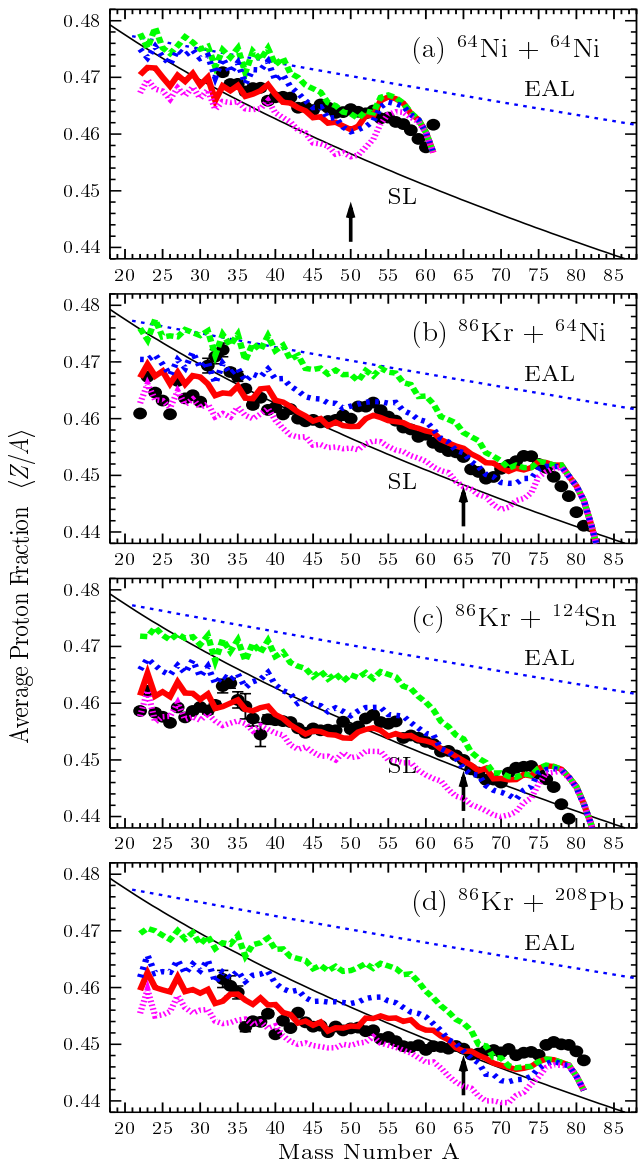}

    \caption{(Color Online)
           Average Z/A vs A.
           (a) $^{64}$Ni\,(25\,MeV/nucleon)+$^{64}$Ni (BigSol data).
           (b), (c), (d) $^{86}$Kr\,(25\,MeV/nucleon) + $^{64}$Ni, $^{124}$Sn, $^{208}$Pb,
           respectively (MARS data).
           Solid points: data.
           Thin solid line (SL): line of stability. Thin
           dotted line (EAL): evaporation attractor line.
           Thick lines: DIT/SMM05 calculations. Dotted lines,
           top to bottom:  with $C_{\rm sym}$ = 25, 20, 15 MeV, respectively.
           Thick solid line:  with $C_{\rm sym}(E^{*}/A)$ given by
           the thick line in Fig. 2.
           Arrows: multifragmentation vs compound nucleus regime.
          }
    \label{za}
    \end{figure}

Fig. 1 shows the average proton fraction $\langle Z/A \rangle$ of
the fragments with respect to mass for the reactions presented in
order of increasing system $N/Z$. The experimental data are shown by
solid symbols and span the whole range of masses from evaporation
residues down to heavy IMF. Vertical arrows indicate an approximate
separation between  heavy residues from evaporation and from the
onset of multifragmentation.
The thin solid line (marked ``SL'') gives the line of $\beta$
stability  and the thin dotted line (marked ``EAL'') represents the
evaporation attractor line \cite{EAL} (corresponding to the locus of
fragment yields produced by evaporation of neutrons and charged
particles from highly excited nuclei close to stability).
The evaporation residue regime is populated mostly  by neutron
deficient products, except for the  heaviest  fragments from
$^{86}$Kr+$^{64}$Ni and $^{86}$Kr+$^{124}$Sn corresponding to very
peripheral products \cite{GSprl}. Fragments on the left side of the
arrow (produced above the multifragmentation threshold of $E^{*}/A
\sim $ 2 MeV) show progressively smaller values of $Z/A$ (Fig. 1)
and cross the line of  $\beta$ stability.

The results of the DIT/SMM05 calculations are presented by thick
lines. The dotted lines correspond to calculations with constant
values of the symmetry energy of $C_{\rm sym}$ = 25, 20, 15 MeV from
the upper to the lower line, respectively. We observe that the
standard calculation with $C_{\rm sym}$ = 25 MeV produces fragments
that are, on average, more neutron deficient than the observed
fragments. With $C_{\rm sym}$ = 20 MeV, the agreement with the data
seems to improve, except for the lowest masses, whereas with $C_{\rm
sym}$ = 15 MeV, only the latter group of masses is described.
We point out that the measured heavy fragment data at 25 MeV/nucleon
span a continuous range of excitation energies up to $E^{*}/A$ =
4.0--4.5 MeV (corresponding to $\sim$ 95\%
of all produced fragments). 
Assuming a continuous behavior of the symmetry energy from low to
high excitation, we tested various forms of the dependence of
$C_{\rm sym}$ on excitation energy $E^{*}/A$. We found that the form
represented by the thick solid line in Fig. 2, referred to as form
``2--4'' in the following, [assigning $C_{\rm sym}$ = 15 MeV to
events with $E^{*}/A$ $>$ 4 MeV] provides a remarkable agreement of
the calculations with the data from multifragmentation-like
processes, as shown by the solid lines in Fig. 1.

The thin solid lines and the thin dashed line in Fig. 2 indicate
three additional forms of $C_{\rm sym}(E^{*}/A)$ that we tested. The
results of $\langle Z/A \rangle$ for Kr+Sn are shown in Fig. 3 by
the corresponding thin lines (the lower one corresponding to the
left thin line of Fig. 2 and the upper one to the right thin line).
In Fig. 3, we notice that the thin solid lines embrace the thick
line obtained with the ``2--4'' form of $C_{\rm sym}(E^{*}/A)$,
whereas the thin dashed line is almost indistinguishable from the
thick one. Similar results are obtained for the other reactions of
this work.
Furthermore, assuming that the multifragmentation threshold is at
$E^{*}/A$ = 2 MeV, we may only adopt forms of $C_{\rm sym}(E^{*}/A)$
with the condition $C_{\rm sym}$ = 25 MeV at the threshold and
below. An appropriate such form appears to be the``2--4'' form
represented by the thick solid line in Fig. 2, as mentioned
previously.
Consequently, within the present SMM interpretation of fragment
formation, the symmetry energy coefficient appears to
decrease 
from 25 MeV in the compound nucleus regime  (left arrow in Fig. 2)
to 15 MeV  in the bulk multifragmentation regime (right arrow in
Fig. 2). This phenomenological evolution of the symmetry energy of
hot heavy fragments is the main result of the present study.

For comparison, in Fig. 2 we include the values of the symmetry
energy coefficient obtained in our recent work on heavy fragment
isoscaling from the $^{86}$Kr  (circle, square) and $^{64}$Ni
(triangle) reactions \cite{GS-csym}. We notice that, in spite of the
different approach of symmetry energy determination, an overall
agreement in the trend is observed.
We speculate that the reduction  of the symmetry energy with
excitation observed in this work  and the aforementioned previous
studies \cite{Iglio,Shetty05,LeFevre,Henzlova} may originate from
in-medium modifications of the properties of the hot primary
fragments in their dense surrounding
\cite{Botvina-review,in-medium}. Detailed microscopic calculations
are necessary to shed light on this interesting and challenging
issue \cite{in-medium}.



    \begin{figure}[h]                                        
    \includegraphics[width=0.42\textwidth, height=0.20 \textheight ]{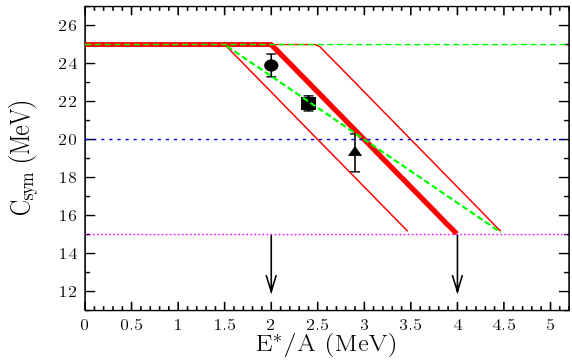}

    \caption{(Color online)
    Thick solid line: form ``2--4'' of $C_{\rm sym}(E^{*}/A)$
    in the SMM05 calculations.
    Thin lines: other forms of $C_{\rm sym}(E^{*}/A)$
    tested (see text and Fig. 3). Thin horizontal lines (top to bottom): $C_{\rm sym}$ =
    25, 20, 15 MeV, respectively,
    used in Fig. 1 calculations.
    Solid points: values of $C_{\rm sym}$ from \cite{GS-csym}.
            }
    \label{csym}
    \end{figure}


    \begin{figure}[h]                                        
    \includegraphics[width=0.47 \textwidth, height=0.20 \textheight ]{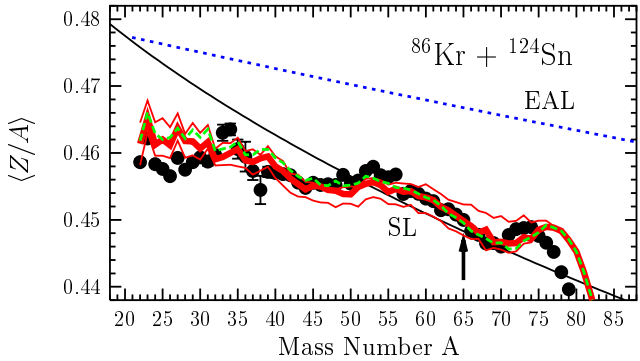}

    \caption{(Color Online)
           $\langle Z/A \rangle$ vs A for $^{86}$Kr   +   $^{124}$Sn.     
           Solid points (data), lines SL, EAL and arrow as in Fig. 1.
           Thin lines tracing the data: DIT/SMM05 calculations with $C_{\rm sym}(E^{*}/A)$
           forms
           presented by the respective lines in Fig. 2 (see
           text).
          }
    \label{za}
    \end{figure}


    \begin{figure}[h]                                        
    \includegraphics[width=0.47\textwidth, height=0.50\textheight ]{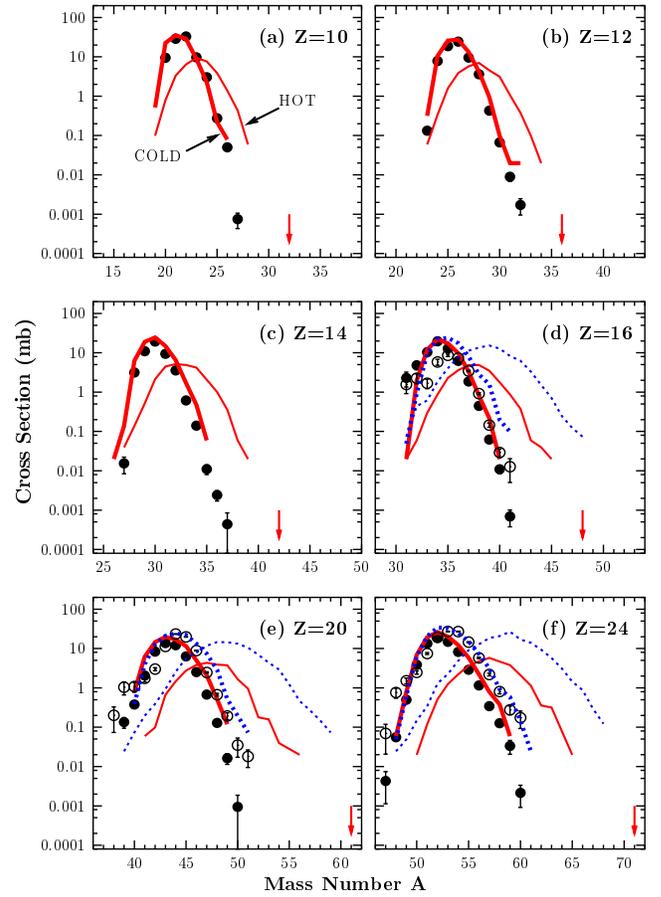}

    \caption{(Color online)
           Comparison  of experimental mass distributions 
           with DIT/SMM05 calculations.
           Solid points: $^{86}$Kr+$^{64}$Ni data.
           Open points:  $^{86}$Kr+$^{208}$Pb data.
           Full lines: calculations for
           $^{86}$Kr+$^{64}$Ni; thick: final (cold) fragments, thin:
           primary (hot) fragments.
           Dotted lines: calculations for
           $^{86}$Kr+$^{208}$Pb; thick: final (cold) fragments, thin:
           primary (hot) fragments. Arrows: neutron-drip
           line \cite{Moller}.
           }
    \label{yield}
    \end{figure}


Apart from the average $Z/A$ properties, the present calculations
describe successfully the characteristics of the observed fragments
including the widths of the isotopic mass distributions, the
velocity distributions and the production cross sections. A full
report on these results is currently underway.
For completeness, we present in Fig. 4 a comparison of the
experimental mass distributions for several elements
with the DIT/SMM05 calculations for two reactions:
$^{86}$Kr+$^{64}$Ni and $^{86}$Kr+$^{208}$Pb. The ``2--4" form of
$C_{\rm sym}(E^*/A)$ dependence (thick line in Fig. 2) was used in
these calculations.
For Kr+Ni, the data 
 are shown by the full circles and the calculations by the thick solid
lines (the thin solid lines giving the hot primary fragment
distributions for these elements). Similarly, for Kr+Pb, the data
are given by open circles and the calculations by thick dotted lines
(again, the thin dotted lines describing hot fragments).
We find an overall satisfactory agreement of the calculations with
the mass distribution data. Interestingly, the Kr+Pb data extend to
more neutron-rich products than those of Kr+Ni (demonstrating the
significant role of the target $N/Z$ at this energy regime
\cite{GSprl,GS-plb04}),
a trend well described by the calculation.
Finally, we point out that the distributions of the hot  primary
fragments are wide and include very neutron-rich nuclei.
Particularly, for the Kr+Pb reaction (the most neutron-rich system
studied), they are very broad and  extend towards the neutron
drip-line \cite{Moller} (arrows in Fig. 4).
Our calculations indicate that the distributions of these hot exotic
nuclei are sensitive to their masses and, in turn, sensitive to the
evolution of the symmetry energy with excitation.

We wish to emphasize that high-resolution mass spectrometric data on
heavy fragment (heavy residue and heavy IMF) distributions --
spanning the full spectrum of reaction mechanisms (and excitation
energies) from the traditional compound nucleus towards
multifragmentation -- may provide a promising experimental probe of
the properties of hot and extremely neutron-rich fragments  when
coupled with detailed calculations as presented in this study.  From
an experimental point of view, we wish to propose systematic
measurements on residue distributions on a variety of neutron-rich
systems. The energy regime of $\sim$25\,MeV/nucleon has two main
advantages: (a) $N/Z$ equilibration is attained (e.g.
\cite{GS-plb04}), thus the target has maximum contribution to the
$N/Z$ distribution of the quasiprojectiles and (b) the excitation
energy is not too high ($E^{*}/A$ $<$ 4 MeV), so that hot heavy
primary fragments are abundantly present.
A very exciting extension of these studies in the future involves
reactions with very neutron-rich rare beams on heavy neutron-rich
targets 
in current or planned radioactive beam facilities (e.g. the Advanced
Exotic Beam facility in the US). 

We believe that comparisons of high-resolution heavy fragment data
with precise model calculations 
can provide valuable information on the distributions and the
properties of hot and very neutron-rich nuclei towards and beyond
the neutron drip-line.
Knowledge on the properties of exotic nuclei produced in hot and
dense environments are essential for astrophysical calculations,
such as the composition and dynamics of core-collapse supernova and
the course of the relevant nucleosynthesis processes (e.g., the r
process).

%
%



In summary, experimental data on isotopic distributions of
projectile-like fragments from peripheral and semiperipheral
collisions of 25\,MeV/nucleon $^{86}$Kr and $^{64}$Ni beams on heavy
targets are compared to calculations involving the Statistical
Multifragmentation Model (SMM). Within the adopted SMM picture, the
present study suggests a gradual decrease of the symmetry energy
coefficient of the hot primary fragments from 25 MeV  in the
compound nucleus regime towards 15 MeV in the multifragmentation
regime.
The isotopic distributions of the hot fragments following the
multifragmentation stage are very broad and, for the most
neutron-rich systems, are found to extend to exotic neutron-rich
nuclei. These findings are of importance in calculations of the
isotopic composition in hot and dense astrophysical environments,
such as those encountered in supernova.




We are thankful to L. Tassan-Got for the DIT code. We gratefully
acknowledge the support of the Cyclotron Institute staff during the
measurements. Financial support for this work was provided, in part,
by the U.S. Department of Energy under Grant No. DE-FG03-93ER40773
and by the R.A. Welch Foundation under Grant No. A-1266. A.S.B was
supported in part through grant RFFR 05-02-04013 (Russia) and M.V.
through VEGA-2/5098/25 (Slovakia).


\bibliography{nz_paper}



\end{document}